# On the Possibility of an Artificial Origin for `Oumuamua

Avi Loeb

Astronomy Department, Harvard University, 60 Garden Street, Cambridge, MA 02138, USA

**Abstract**

The first large interstellar object discovered near Earth by the Pan STARRS telescope, `Oumuamua, showed half a dozen anomalies relative to comets or asteroids in the Solar system. All natural-origin interpretations of `Oumuamua's anomalies contemplated objects of a type never-seen-before, such as: a porous cloud of dust particles, a tidal disruption fragment or exotic icebergs made of pure hydrogen or pure nitrogen. Each of these natural-origin models has major quantitative shortcomings, and so the possibility of an artificial origin for `Oumuamua must be considered. `Oumuamua's anomalies suggest that it might have been a thin craft - with a large area per unit mass - pushed by the reflection of sunlight; sharing qualities with the thin artifact 2020 SO - launched by NASA in 1966 and discovered by Pan STARRS in 2020 to exhibit a push away from the Sun with no cometary tail. The *Galileo Project* aims to collect new data that will identify the nature of `Oumuamua-like objects in the coming years.

## I. Anomalies

On October 19, 2017, the first large object from outside the Solar system was discovered near Earth by the Pan STARRS telescope in Hawaii. It was given the name, `Oumuamua, which means `a messenger from afar, arriving first' in the Hawaiian language (Meech et al., 2017; Bannister et al., 2020; see Figure 1). This was unexpected, and indeed, a decade ago Moro-Martin, Turner & Loeb (2009) forecasted that Pan STARRS would not detect any rock from other stars based on what we know about the Solar system.

The second interstellar object was discovered named 2I/Borisov after its discoverer, appeared to be a familiar comet (Krishnakumar et al., 2022; Gibbs, 2019). Its ordinary properties only highlight how unusual `Oumuamua was.

Astronomers initially assumed that `Oumuamua is a comet, because comets are most loosely bound to the Sun, residing in the Oort cloud at the periphery of the Solar system where they can be easily sent to interstellar space by the perturbation of a passing star. But there was no visible cometary tail around `Oumuamua. Moreover, a cometary origin implies that `Oumuamua would have inherited the motion of its parent star. But instead it was found to originate from the so-called "Local Standard of Rest" (see Figure 2). This frame averages over the motions of all stars near the Sun, and only 1 in 500 stars is so much at rest as `Oumuamua was in that frame, before the Sun's gravitational force deflected its trajectory (Mamajek, 2017; see Figure 3). The initial velocity of `Oumuamua was unusual for a natural object that originated from a nearby star. If it were a comet originating from the outskirts of another planetary system, it would have inherited the motion of its host star. But since fewer than 0.2% of all stars share `Oumuamua's kinematic origin at the Local Standard of Rest, it was unlikely to originate from a nearby star system. And distant stars are moving much faster relative to this local frame because of the Galactic rotation velocity profile. If `Oumuamua is artificial, then its initial conditions might have been tailored, for example to hide the identity of its host star system.

These were just the initial anomalies that made `Oumuamua different from all the comets and asteroids that we had seen before in the Solar system. As it tumbled every eight hours (see Figure 4), the brightness of sunlight reflected from it changed by a factor of ten. This meant that it has an extreme shape, which at the ~90% confidence level was disk-like (Mashchenko, 2019). The Spitzer Space Telescope did not detect any carbon-based molecules or dust around `Oumuamua, setting a tight limit on ordinary cometary activity (Trilling et al., 2018). The lack of heat, detectable in the infrared, placed an upper limit of about 200 meters on its size, the scale of a football field. But most remarkably, `Oumuamua exhibited an excess push away from the Sun which would have required it to lose ~10% of its mass if it was caused by the rocket effect from normal cometary evaporation (Micheli et al., 2018). An extensive evaporation of this magnitude was absolutely ruled out by the Spitzer telescope data; moreover, the repulsive force declined smoothly with distance from the Sun, showing no change in spin or sudden kicks as routinely observed from localized jets on the surface of comets (Rafikov, 2018). Finally, there was no apparent cut-off in the push at the distance beyond which evaporation of water ice by the heating of sunlight is expected to stop (see Figure 5).

The excess force without a cometary tail implied that this object is not a familiar rock. Since the push away from the Sun was consistent with a smooth inverse-square law, I reasoned that it may result from the reflection of sunlight from a thin object (Bialy & Loeb, 2018). For the reflection of sunlight to exert a strong enough force, the object had to be thinner than a millimeter, like a light sail. Since nature does not make thin objects, I suggested that it might be artificial in origin (Loeb, 2018a; 2018b; 2018c).

This possibility should encourage scientists to obtain better data on an `Oumuamua-like object in the future. This could include spectroscopy with the James Webb Space Telescope (Forbes & Loeb, 2019) or a high-resolution image of `Oumuamua-like objects with a flyby mission that will rendezvous with such an object along its approach towards Earth, as planned in the new Galileo Project (GP, 2021). An object the size of `Oumuamua at the Earth-Sun separation occupies an angular scale of 100 micro-arcseconds, and to resolve it in visible light requires a kilometer-size aperture, two orders of magnitude larger than the biggest ground-based telescopes on Earth.

In September 2020, another object (TTSO, 2020) was discovered by Pan STARRS, sharing `Oumuamua's anomalies of no cometary outgassing and excess push away from the Sun as a result of the reflection of sunlight. It was given the astronomical name 2020 SO and later found to be a thin shell of a rocket booster from a 1966 NASA mission to the Moon. It had thin walls and hence a large area for its mass. It was not designed to be a light sail but was thin for a different purpose. Its discovery illustrates that the difference between a typical rock and an object with an area to mass ratio large enough to imply artificial construction can be inferred from the unusual dynamics of the object. We know that humanity manufactured 2020 SO. The question is who manufactured `Oumuamua?

An anthropologist finding a curious rock and wondering whether it is an artifact would first suggest that it is a rock of a new type, in the same way that earthlings who studied the anomalies of the first large interstellar object `Oumuamua suggested that it is a comet of a type "never seen before", such as an iceberg made of pure hydrogen or pure nitrogen, even though these possibilities face "serious difficulties" in the words of some of their proponents (Levine et al. 2021). Extraordinary evidence requires extraordinary funding in order to substantiate it. Mainstream scientists should allocate funds for finding the next `Oumuamua-like object and identifying beyond a reasonable doubt its nature, just as the OSIRIS-Rex mission identified the nature of the asteroid Bennu on which it had landed.

## II. Possible Natural Origins and Their Challenges

Astronomers who attempted to explain the anomalies of `Oumuamua by a natural origin were all forced to contemplate objects that were never seen before, with major quantitative challenges. These possibilities are:

(i) a porous structure with a mean density a hundred times lower than air (Moro-Martin, 2019; Luu et al., 2020) - which is unlikely to maintain its integrity after being heated to hundreds of degrees by the Sun (Forbes & Loeb 2019);
(ii) fragments from tidal disruption (Zhang & Lin 2020) – whose shape is more likely to be that of a cigar than a disk as inferred for `Oumuamua (Mashchenko, 2019);
(iii) an iceberg of molecular hydrogen (Seligman & Laughlin 2020) – which evaporates too quickly along its interstellar journey and must have been unreasonably massive initially in order for it to survive the trip from the distance of the nearest molecular clouds (Hoang & Loeb 2020; see Figure 6);
(iv) a nitrogen iceberg chipped off the surface of a planet like Pluto around another star (Desch & Jackson 2021)– a mechanism that cannot supply enough material to explain the implied abundance of objects like `Oumuamua (Levine et al., 2021; Siraj & Loeb, 2021; Phan et al., 2021; see Figure 7).

Cases (iii) and (iv) require new nurseries of interstellar objects of exotic composition that was never previously observed. Since the first large interstellar object was found to be anomalous, there must be abundant source never imagined before. A confirmation of any of them by future data on `Oumuamua-like objects, like compositional analysis from spectroscopy or high-resolution imaging, would imply new insights on the sources of interstellar objects and is therefore worth any effort to uncover with certainty.

## III. The Possibility of an Artificial Origin

Given these challenges to natural origins of `Oumuamua and the similarity in the anomalous dynamics of 2020 SO and `Oumuamua, the possibility of an artificial origin should be considered (see Figure 8).

Thanks to the generous donations from people who were inspired by the vision of my book *Extraterrestrial*, I was able to inaugurate in July 2021 the ***Galileo Project*** (GP, 2021). One of the major goals of the project is to search for `Oumuamua-like objects in future surveys, like with the upcoming Vera Rubin Observatory. An early alert to `Oumuamua-like objects would allow the design of a space mission that will intercept their trajectories and take close-up photographs of them. Such data could resolve their nature and unambiguously determine whether they are natural or artificial in origin.

The Galileo Project, announced on July 26, 2021, pioneers a search for extraterrestrial equipment near Earth. It has two branches: the first aiming to identify the nature of interstellar objects that do not resemble comets or asteroids, like `Oumuamua, and the second to understand objects closer to Earth, similar to those mentioned in a recent report from the Director of National Intelligence to the US Congress. By now, the Galileo research team includes more than a hundred scientists who plan to design a space mission to take a high-resolution photograph of the next `Oumuamua through a rendezvous with its trajectory on its approach to Earth - following an alert from the Vera Rubin Observatory, similarly to past mission designs (Snodgrass & Jones, 2019; Hein et al., 2021). The project will also assemble its first custom built telescope system on the roof of the Harvard College Observatory. The system will record continuous video and audio of the entire sky, and track objects of interest – with the video spanning visible, infrared and radio bands. Artificial intelligence algorithms will distinguish birds from drones, airplanes or something else. Once the first system will operate successfully, the Galileo Project will make copies of it and distribute them in many geographical locations.

The Galileo Project has drawn a remarkable base of expert volunteers, from astrophysicists and other scientific researchers, to hardware and software engineers, to non-science investigators and generalists who volunteer their time and effort to the project in various ways. The project brings together a broad community of members, including believers and skeptics, united by the pursuit of evidence through new telescopes without prejudice. The project values the input of many different voices since its main conclusions will be guided by evidence. The rapid progress it already made is a testament to its open approach. As different as the perspectives of the researchers and affiliates may be, however, every contributor to the Galileo Project is bound by the following ground rules:

1. The Galileo Project is only interested in openly available scientific data and a transparent analysis of it. Thus, classified information, which cannot be shared with all scientists, cannot be used. Such information would compromise the scope of our scientific research program, which is designed to acquire valid scientific data and provide transparent analysis of this data.

2. The analysis of the data will be based on known physics and will not entertain ideas about extensions to the standard model of physics. The data will be published and available for peer review as well as to the public, when such information is ready to be made available. The scope of the research efforts will always remain in the realm of scientific hypotheses, tested through rigorous data collection and sound analysis.

The Galileo team developed a design of telescope systems optimized for imaging unidentified objects near Earth, as well as a blueprint for a space mission to image unusual interstellar objects like `Oumuamua, to be launched within a few years.

Artificial Intelligence (AI) algorithms can distinguish extraterrestrial equipment from familiar objects like a meteor, or an atmospheric phenomenon. Since there are no birds, airplanes or lightning above the Earth's atmosphere, any object with an elevation larger than 50 kilometers would appear highly unusual, as long as it is not a rare meteorite.

Finding equipment from an extraterrestrial technological civilization would have a major impact on the future of humanity. Here's hoping that we will be open minded enough to search for objects that resemble the equipment that our technological civilization is launching to space. We know that half of the Sun-like stars host a planet the size of the Earth roughly at the same separation (Bryson et al., 2021). Many of these stars formed billions of years before the Sun, allowing for the possibility that numerous probes were sent to interstellar space. Ridiculing the notion that `Oumuamua may have been artificial in origin will not get rid of our neighbors (Loeb 2021). As Galileo Galilei instructed us four centuries ago, the nature of celestial objects must be found through our telescopes rather than philosophical prejudice. The nature of `Oumuamua is not a question that will be settled by philosophical arguments (Cowie, 2021). Instead, it is a scientific question which can be clarified with better data.

**FIGURES**

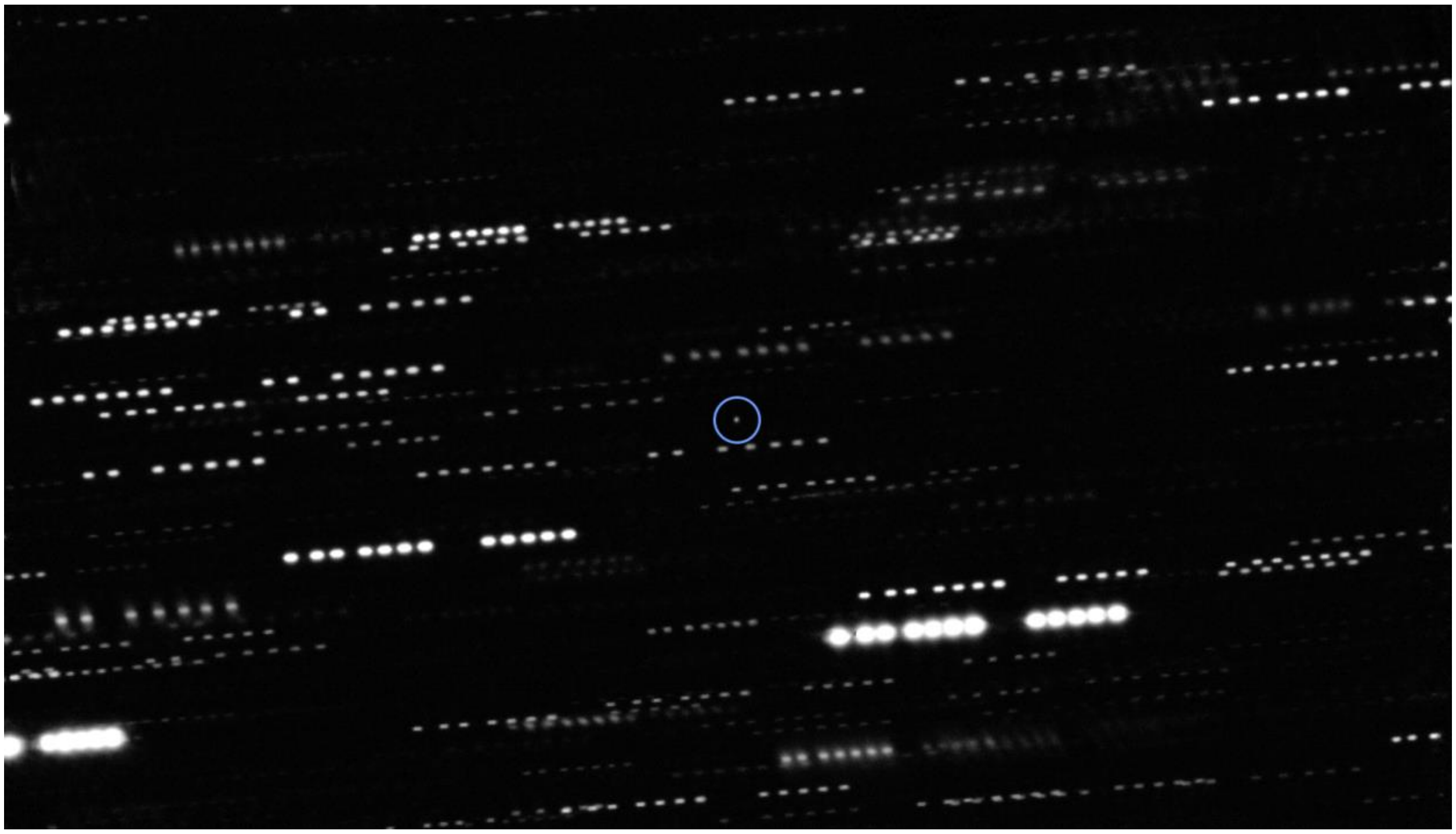

**Figure 1:** Combined telescope image of the first interstellar object `Oumuamua, circled in blue as an unresolved point source at the center. It is surrounded by the trails of faint stars, each smeared into a series of dots as the telescope snapshots tracked the moving `Oumuamua. (Credit: ESO/K. Meech et al.)

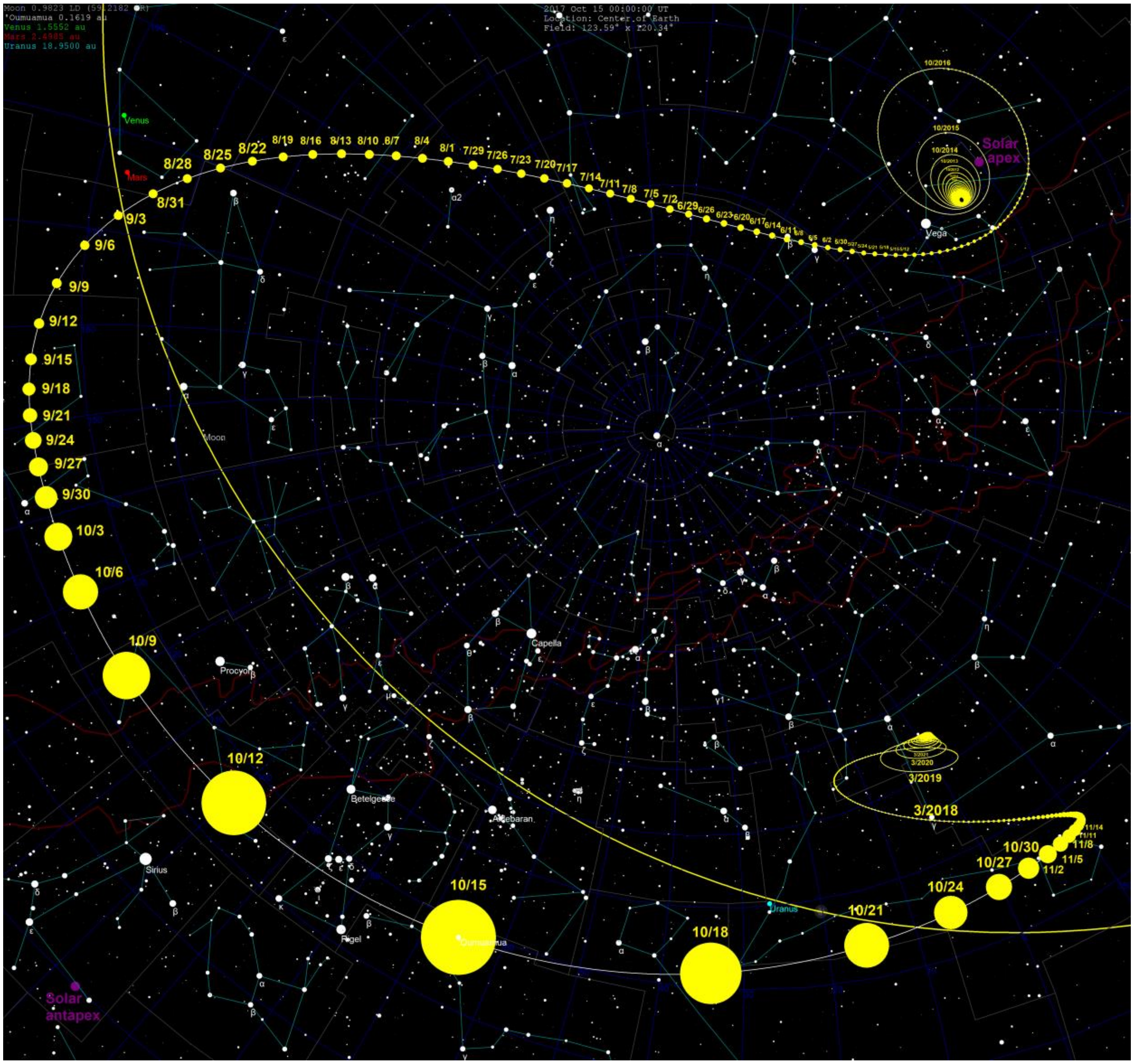

**Figure 2:** Sky path of `Oumuamua, labeled by date, as seen from Earth. The relative size of each circle gives a sense of the changing distance of `Oumuamua along its apparent trajectory. Also shown are the direction of motion of the Sun in the Local Standard of Rest (purple, labeled “Solar apex”), Venus (green), Mars (red) Uranus (turquoise) and the opposite direction to the motion of the Sun (purple, labeled “Solar antapex”). `Oumuamua’s trajectory moved from the Local Standard of Rest to south of the ecliptic plane (marked by the thin yellow line) of the Solar System between September 2 and October 22, 2017. (Credit: JPL Horizons)

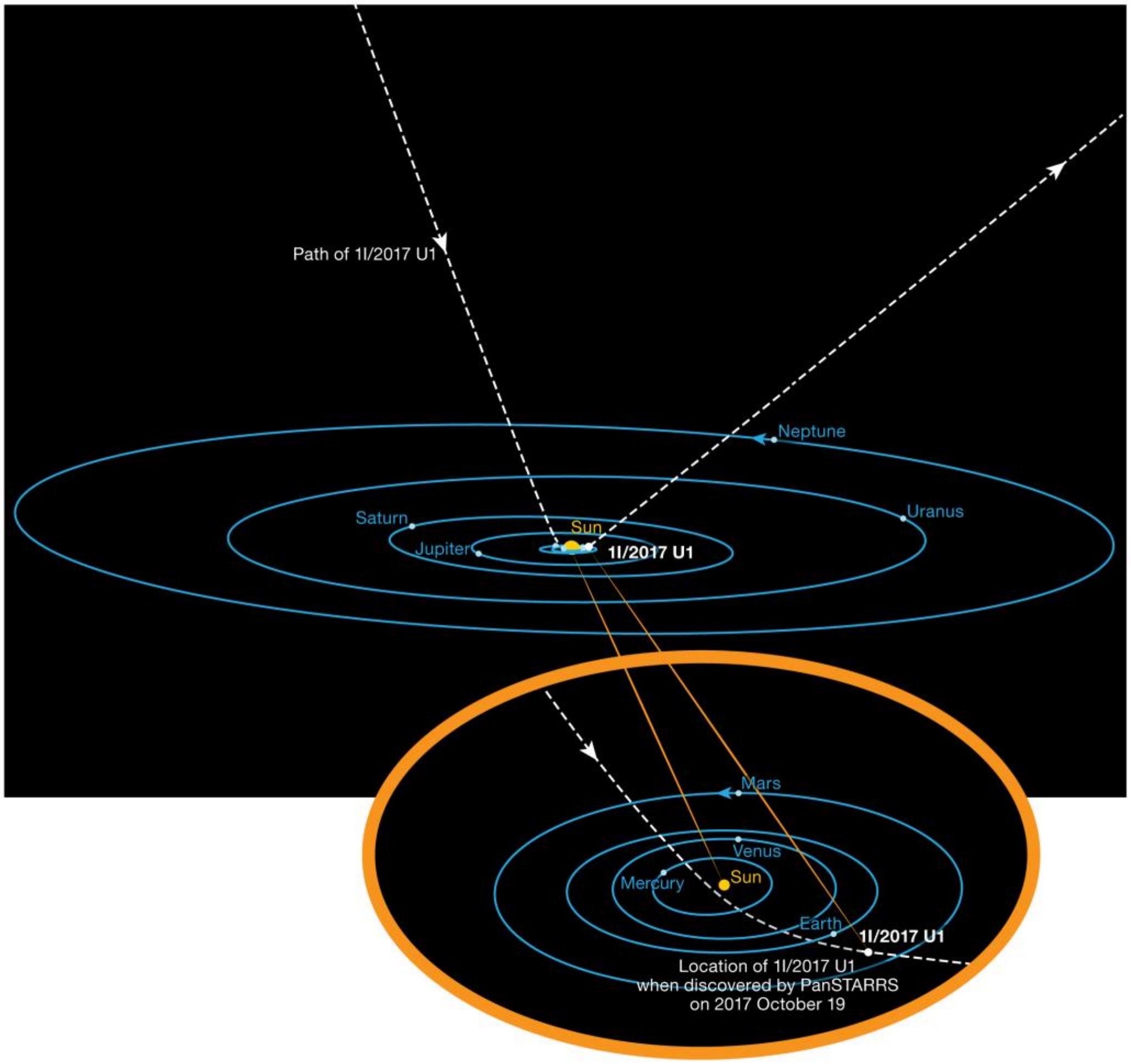

**Figure 3:** Trajectory of \`Oumuamua through the Solar System. Unlike all asteroids or comets observed before, this orbit is not bound by the Sun's gravity. \`Oumuamua originated from interstellar space and will return there with a velocity change as a result of its passage near the Sun. The velocity of \`Oumuamua relative to the Local Standard of Rest was changed in magnitude by the gravitational force of the Sun. Its hyperbolic orbit was inclined relative to the ecliptic plane of the Solar System and did not pass close to any of the planets on the way in. (Credit: ESO/K. Meech et al.).

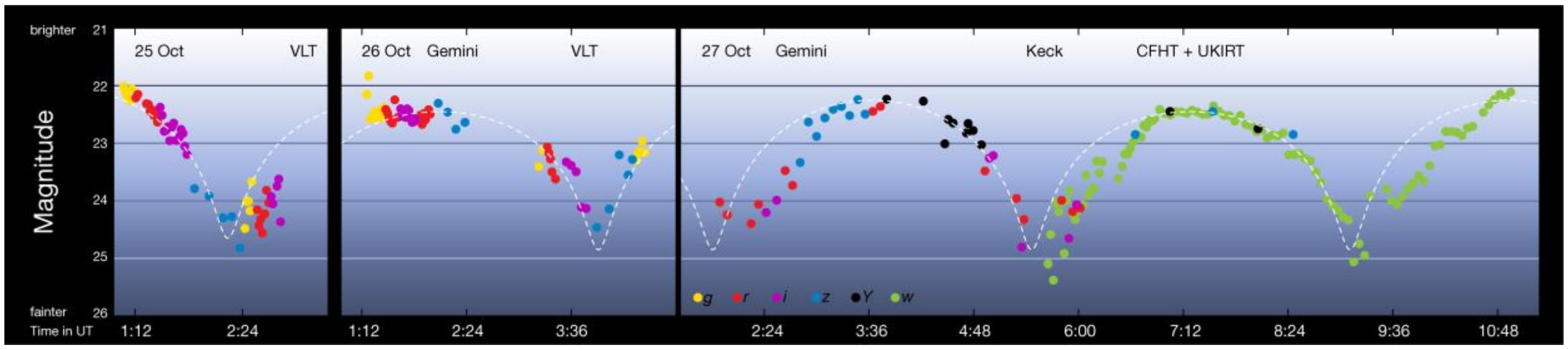

**Figure 4:** Variation in brightness of `Oumuamua as observed by various telescopes during three days in October 2017. Different colored dots represent measurements through different filters in the visible and near-infrared bands of the color spectrum. The amount of reflected sunlight changed periodically by about a factor of ten (2.5 magnitudes) as `Oumuamua rotated every 8 hours. This implied that it has an extreme shape which is at least ten times longer than it is wide when projected on the sky. The dashed white line shows the curve expected if `Oumuamua were an ellipsoid with a 1:10 aspect ratio. However, the best fit to the light curve from its tumbling motion implies a flattened, pancake-shaped configuration rather than an oblong, cigar-shaped object as commonly depicted in the media. (Credit: ESO/K. Meech et al.)

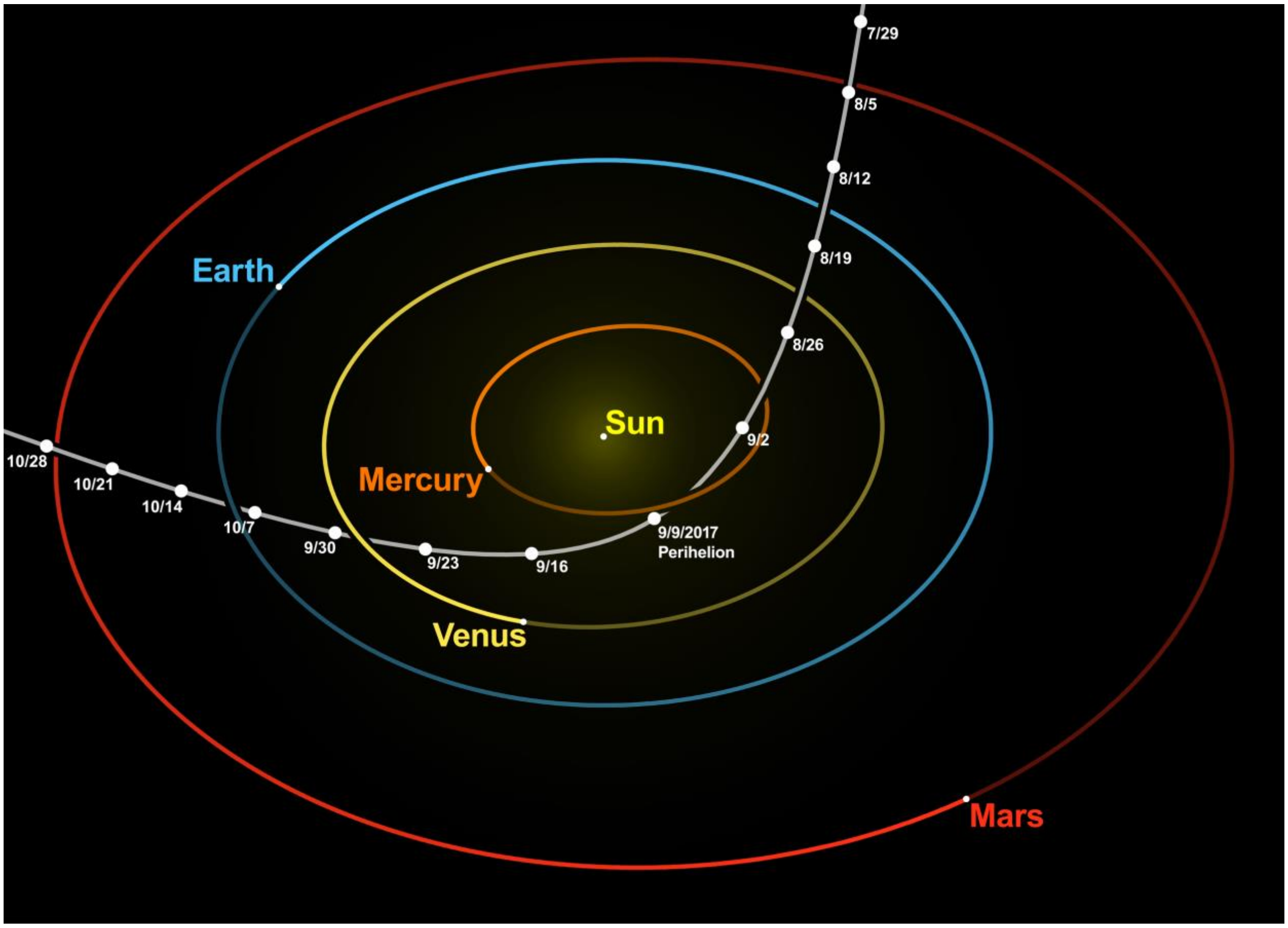

**Figure 5:** Trajectory of `Oumuamua through the inner region of the Solar System, dated weekly. The planet positions are fixed at the time of `Oumuamua's closest approach to the Sun (perihelion) on September 9, 2017.

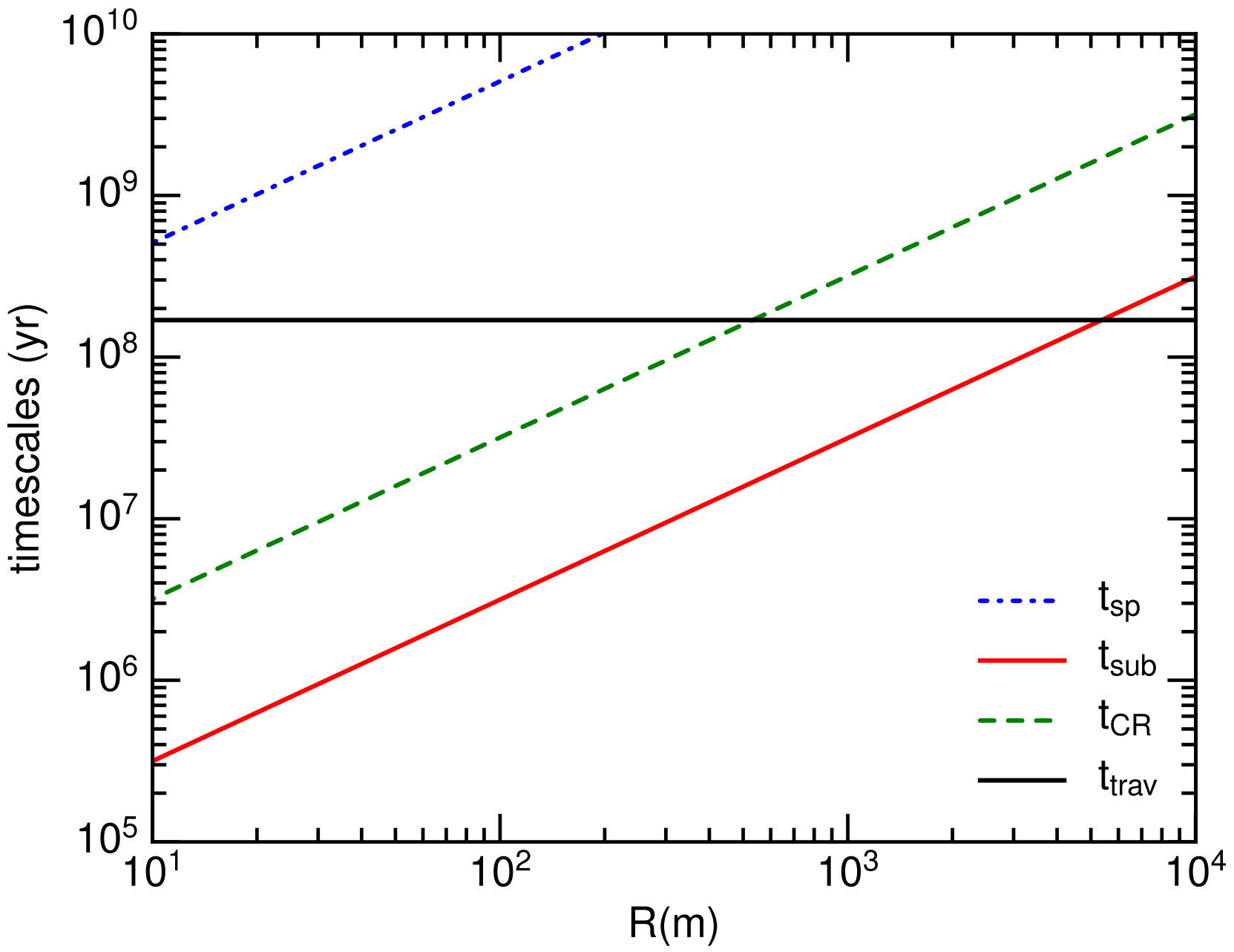

**Figure 6:** Comparison of various destruction timescales for a hydrogen $H_2$ iceberg (slanted colored lines) as a function of the object radius (in meters) to the travel time from the likely source of a giant molecular cloud at a distance of 5.2 kpc, assuming a characteristic speed of 30 km/s (horizontal black line). (Credit: Hoang & Loeb 2020)

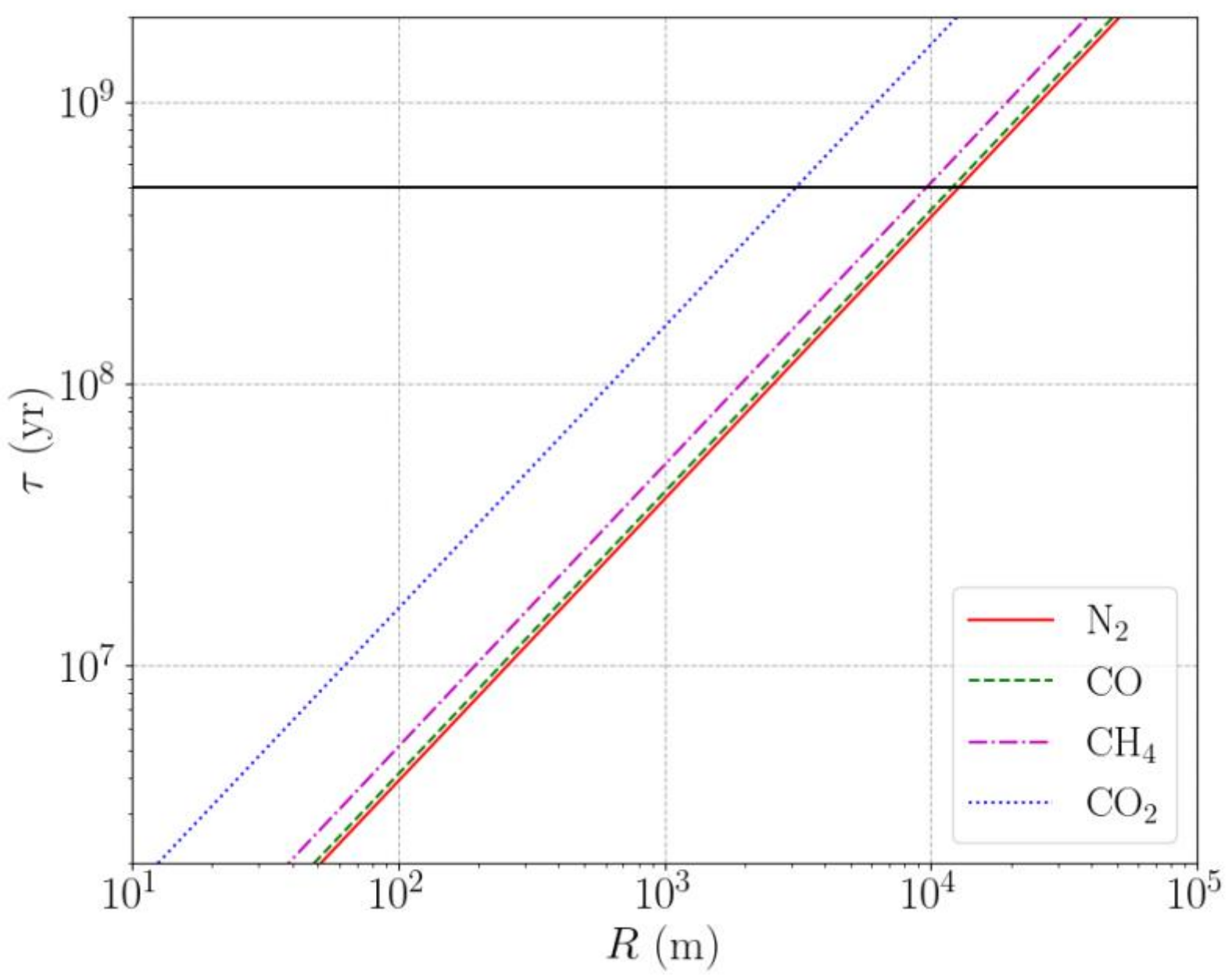

**Figure 7:** Erosion time by cosmic rays for various types of ices including nitrogen N2 (solid red line), CO (dashed green line), CO2 (dotted blue line), and CH4 (dash-dot magenta line) in comparison with the suggested travel time of ~0.5 Gyr for 'Oumuamua (solid black line). A short travel time, would imply origin from nearby young stars, which are much less abundant than old stars. This makes the required nitrogen mass budget untenable. (Credit: Phan, Hoang & Loeb 2021)

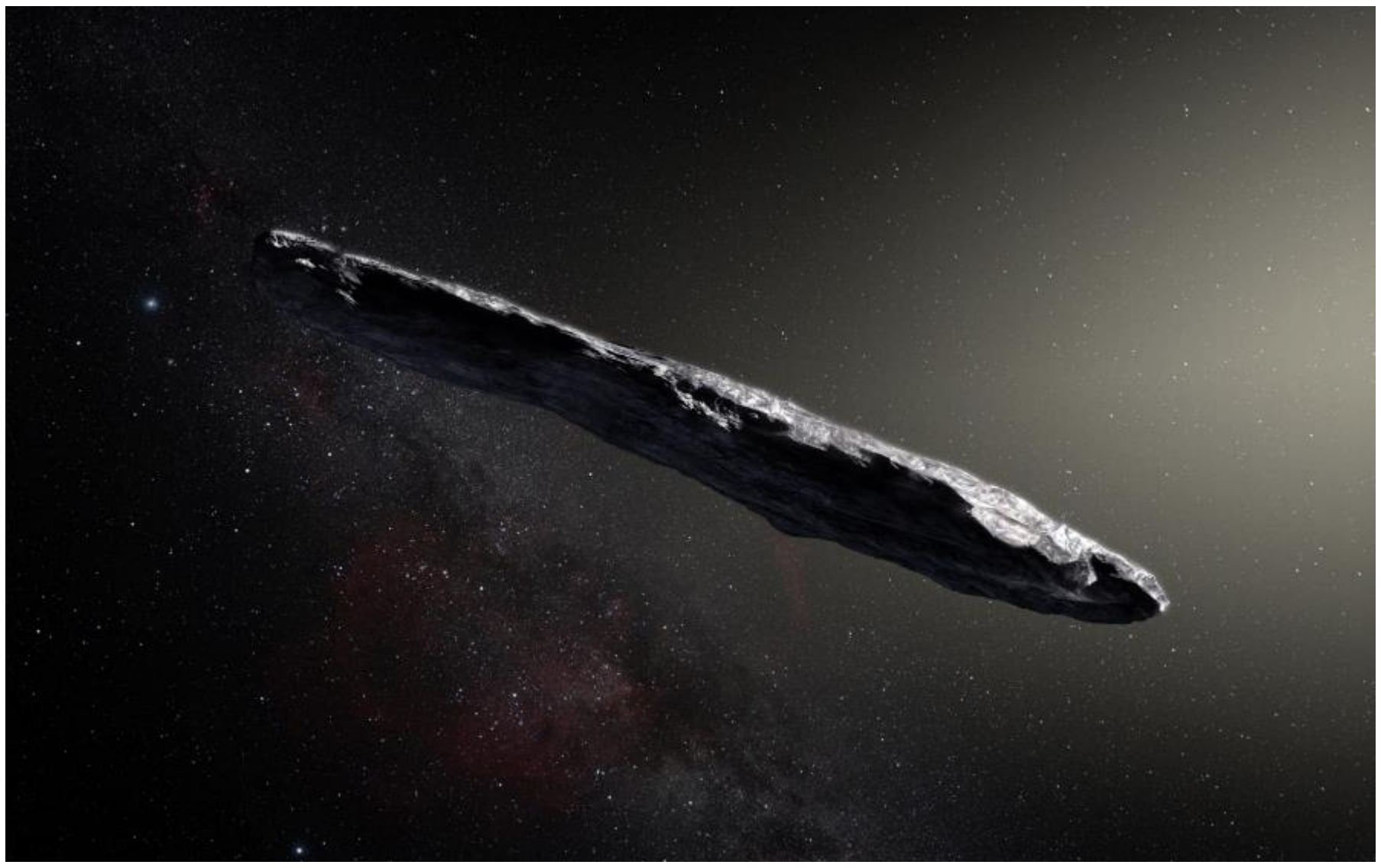

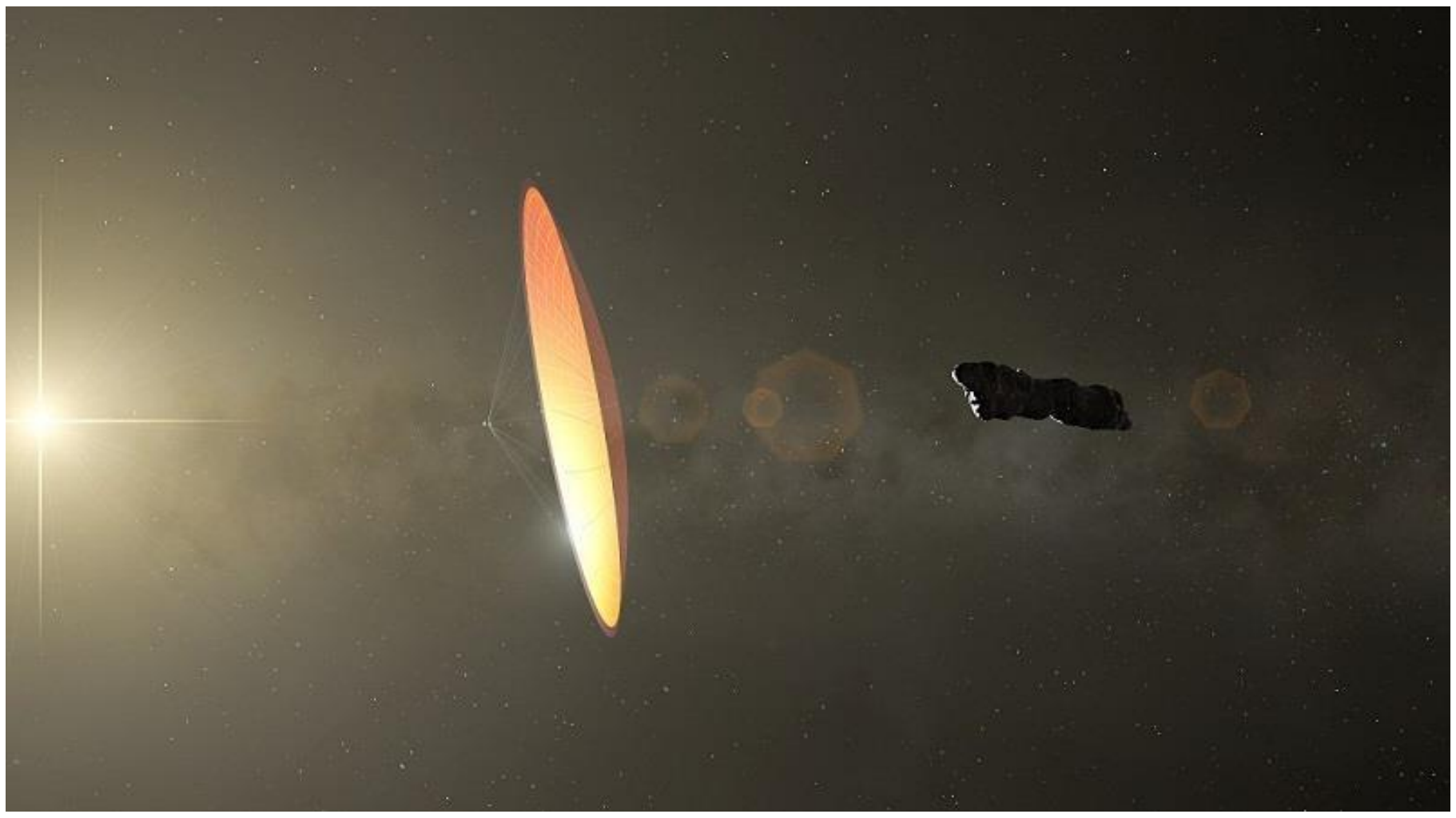

**Figure 8:** Artist's impressions of two possible shapes for `Oumuamua. The object's length is estimated to be between tens to hundreds of meters, up to the size of a football field. It is either an oblong, cigar-shaped rock - as depicted in the upper image (Credit: ESO/M. Kornmesser), or a flattened, pancake-shaped object - as shown in the lower image, which also shows a rock – as a representation of the alternative interpretation (Credit: Mark Garlick). The pancake shape provides the best fit to `Oumuamua's light curve[3]. Even a razor-thin object, like a flat sheet of paper, would appear to possess some width when projected at a random orientation on the sky, so the intrinsic aspect ratio of `Oumuamua can be much smaller than the value of 1:10 inferred from its light curve.